\documentclass[10pt]{article}
\usepackage{amsmath,amssymb}

\usepackage{changepage}

\usepackage[utf8x]{inputenc}

\usepackage{textcomp,marvosym}

\usepackage{cite}

\usepackage{nameref,hyperref}

\usepackage[right]{lineno}

\usepackage{microtype}
\DisableLigatures[f]{encoding = *, family = * }

\usepackage[table]{xcolor}

\usepackage{array}

\newcolumntype{+}{!{\vrule width 2pt}}

\newlength\savedwidth

\usepackage[aboveskip=1pt,labelfont=bf,labelsep=period,justification=raggedright,singlelinecheck=off]{caption}

\makeatletter
\renewcommand{\@biblabel}[1]{\quad#1.}
\makeatother

\usepackage{lastpage,fancyhdr,graphicx}
\usepackage{epstopdf}
\title{A comparison of simple models for urban morphogenesis}
\date{}

\author{
Juste Raimbault\textsuperscript{1,2,3,*}
\bigskip\\
\textsuperscript{1} Center for Advanced Spatial Analysis, University College London
\\
\textsuperscript{2} UPS CNRS 3611 ISC-PIF
\\
\textsuperscript{3} UMR CNRS 8504 G{\'e}ographie-cit{\'e}s
\bigskip\\
* juste.raimbault@polytechnique.edu
}

\begin{document}

\maketitle

%\end{flushleft}
%\end{center}
%\section*{Abstract}
\vspace{1cm}
\begin{abstract}
The spatial distribution of population and activities within urban areas, or urban form at the mesoscopic scale, is the outcome of multiple antagonist processes. We propose in this paper to benchmark different models of urban morphogenesis, to systematically compare the urban forms they can produce. Different types of approaches are included, such as a reaction-diffusion model, a gravity-based model, and correlated percolation. Applying a diversity search algorithm, we estimate the feasible space of each model within a space of urban form indicators, in comparison of empirical values for worldwide urban areas. We find a complementarity of the different types of processes, advocating for a plurality of urban models.
\end{abstract}

%\linenumbers

\section*{Introduction}

% urban form and sustainability - urban form at the mesoscopic scale

Understanding the dynamics of cities is an increasing issue for sustainability, since the proportion of the world population expected to live in cities will grow to a large majority in the next decades, and that cities combine both positive and negative externalities on most aspects. Their complexity implies that quantitative and qualitative predictions are not relevant, but planners can \emph{invent future cities} \cite{batty2018inventing}, what requires though a knowledge of key urban processes which can be acted upon. In that context, the growth of \emph{urban form} in its different definition and scales, is essential \cite{williams2000achieving}. Considering urban form at a mesoscopic scale, i.e. roughly the scale of urban areas, it can be understood as the spatial distribution of activities. More particularly the distribution of population density has a strong impact on commuting, energy consumption and emissions \cite{le2012urban}. Being able to link microscopic processes with the growth of different types of urban form is thus important for a long term planning of sustainable urban systems.

% broad overview of approaches: CA, LUTI. why study simple models (include studied models)

Urban modeling at the mesoscopic scale is the subject of diverse approaches and disciplines. Intra-urban urban economic models, building on classic works such as the Alonso-Mills-Muth model or the Fujita-Ogawa model, propose models linking land-use with land and building markets, which are spatially explicit to different degrees \cite{viguie2012trade}. Transportation and Urban Planning also have a long history in urban dynamics models, including Land-use transport interaction models \cite{wegener2004land}. Spatial interaction models can also be used in a similar manner to study urban dynamics and as a by-product urban form \cite{milton2019accelerating}. Cellular automata models of urban growth are also a privileged approach to study the growth of urban form from a data-driven perspective \cite{batty1997cellular}.

At the interface of physics, artificial life and quantitative geography, a few approaches propose simple models to explain the growth of urban form, and generally rely on an unidimensional description of urban form, namely the distribution of population or of the built environment. In that context, the correlated percolation model introduced by \cite{makse1998modeling} was a precursor. Such models can rely on abstract physical processes but also on agent behavior, such as in the Sugarscape model which according to \cite{batty2007cities} can be considered as a model for human settlements. \cite{murcio2015urban} use migration between cities at multiple scales to simulate urban growth. Diffusion-limited aggregation (DLA) is an other approach transferred from physics to urban modeling \cite{batty1989urban} and has shown relevant to reproduce fractal urban structures and urban migration processes \cite{murcio2009colored}. \cite{murcio2013second} combines DLA with percolation to obtain more realistic urban forms. Closer to the idea of urban morphogenesis, \cite{10.1371/journal.pone.0203516} proposes a reaction-diffusion model to capture fundamental urban growth processes. \cite{li2019singularity} describes an urban growth model based on geographical processes, namely an aggregation of population driven by spatial interaction. All these works have in common to model urban growth in synthetic settings, at a mesoscopic scale, considering population distribution only, and in a stylized way. They furthermore consider diverse processes, remaining simple in their structure although they lead to the emergence of a complex behavior. We will in this paper focus on such models, referring to them as \emph{models of urban morphogenesis}.

% urban multi-modeling/multidimensionality: importance of plurality of views: why benchmark

Exhibiting models with a few number of parameters and processes is useful from an explanative viewpoint, when these can reproduce real world configurations. Having multiple concurrent models which include diverse, complementary or contradictory processes, is furthermore useful for the construction of integrated urban theories, since concurrent explanations can be benchmarked, compared and possibly integrated into multi-modeling approaches. This plurality in urban modeling is intrinsic to a literature with multiple disciplines focusing on a same object of study \cite{pumain2020conclusion}.

% research question / contributions

We propose thus in this paper to benchmark several simple models of urban morphogenesis, in order to understand the potentialities of some of these models to exhibit a complex behavior and reproduce existing urban forms, and compare them in a systematic way. More precisely, our contributions are the following: (i) we integrate four different models (correlated percolation, reaction-diffusion, gravity and exponential mixture) into a single software framework; (ii) we compute measures of urban form for urban areas worldwide; (iii) we apply a novelty search algorithm to the models in order to determine their feasible morphological space, and compare these to real urban form values. This contributes to a general understanding of the complementarity of urban models, more particularly for urban morphogenesis at this scale.

% organization

The rest of this paper is organized as follows: we first describe the models benchmarked and the quantitative measures used for urban form; we describe empirical values of urban morphology indicators for urban areas worldwide; these are then compared to the feasible space of each model obtained with a diversity search algorithm. We finally discuss the implications of these results for theories of urban morphogenesis and possible developments.

\section*{Materials and methods}

%\subsection*{Processes}
% kind of systematic review of possibly involved processes?

\subsection*{Urban morphogenesis models}

We study and compare four different models of urban morphogenesis. We consider a population grid of size $N = W\times H$ (not necessarily square), each cell being characterized by its population $P_i$ with $1\leq i \leq N$.

\subsubsection*{Gravity-based model}

% modified: exponent for spatial interaction model; multiple seeds (polycentricity)

Following the so-called ``first law of geography'', entities in space have interaction patterns which can be described with spatial interaction models \cite{fotheringham1989spatial}, including the gravity model. \cite{li2019singularity} proposed an urban growth model including this process within an iterative growth with population aggregation processes, extending a more simple model introduced by \cite{rybski2013distance}. We generalize this model by adding (i) a hierarchy parameter regarding population aggregation and (ii) seeding multiple initial sites to allow the emergence of polycentric urban forms.

Formally, an initial grid is seeded with $P_0^{(G)}$ sites with population 1, randomly selected. Then, iteratively, one unit of population is added to each cell at each time step with a probability proportional to 
\begin{equation}
p_i \propto \frac{\sum_{j\neq i} P_j^{\gamma_P^{(G)}} \cdot d_{ij}^{- \gamma_D^{(G)}}}{\sum d_{ij}^{- \gamma_D^{(G)}}}
\end{equation}

where probabilities are rescaled such that the cell with the larger value has a probability of $g^{(G)}$. This last parameter allows modifying the speed of growth. The model is stopped when a total population $P_M^{(G)}$ is reached.

\subsubsection*{Reaction-diffusion}

% modify: max density? - NO

In his attempt to understand embryogenesis, Alan Turing proposed to use chemical partial differential equations (PDEs) to model morphogenesis, introducing the nowadays famous reaction-diffusion equations \cite{turing1990chemical}. In such systems, chemical substances react together and diffuse in space, leading to the emergence of complex geometrical patterns. The concept of morphogenesis has been since well used in urban studies \cite{raimbault2018co}, but very few models have actually implemented reaction-diffusion equations, \cite{bonin2014modelisation} being a notable exception. \cite{10.1371/journal.pone.0203516} proposes to capture the fundamental processes of agglomeration economies (positive externalities) leading to aggregation and of congestion (negative externalities) leading to sprawl, as an ``aggregation-diffusion'' model of urban morphogenesis. The model yields indeed in certain limits reaction-diffusion PDEs. Formally, starting from an empty grid, $N^{(R)}$ units of population are added at each time step, and attributed independently to cells with a probability proportional to $P_i^{\alpha^{(R)}}$ (probabilities are rescaled to obtain a probability distribution over all cells). Population is then diffused in space $d^{(R)}$ times with a strength $\beta^{(R)}$. The model is stopped when a maximum population $P_M^{(R)}$ is reached.

\subsubsection*{Correlated percolation}

The first two models presented are iterative and can in theory be used dynamically. Other approaches, closer to procedural modeling \cite{parish2001procedural}, do not simulate the progressive growth of population. They can however capture processes at play in the growth of urban form. The correlated percolation model described by \cite{makse1998modeling} integrates for example clustering processes in cities. A method to generate a spatial field exhibiting long range correlations was introduced for problems in physics by \cite{makse1996method}. It is combined to a monocentric density profile in \cite{makse1998modeling} to produce urban forms. In practice, a correlated field $p_i$ is generated by (i) generating a random spatial field; (ii) compute its spatial Fourier transform; (iii) introduce a correlation by multiplying it with a spectral density function with a power-law exponent $\alpha^{(C)}$; (iv) retrieve a long-range correlated spatial field by taking the inverse Fourier transform. This field is combined to a density field $\rho_i$ to determine a binary value for the cell: it is populated if $p_i > \theta_i$ with $\rho_i = \int_{-\infty}^{\theta_i} d\mathbb{P}(p_i)$. We generalize the initial model by taking a polycentric density field with 

\begin{equation}
\rho_i (\vec{x}) = \sum_{j=1}^{n^{(C)}} \exp\left(\lVert\vec{x} - \vec{x}_j\rVert / r_j\right)
\label{eq:kernelmixture}
\end{equation}

where the kernel centers are chosen at random and kernel sizes $r_j$ are taken such that kernel populations follow a rank-size law of hierarchy $\beta^{(C)}$ and the largest kernel has a fixed size $r_0^{(C)}$.

\subsubsection*{Kernel mixtures}

Finally, to provide some kind of null model to understand the advantages of each approach compared to a simple description of population distribution, we also include urban forms generated as kernel mixtures. We consider in particular exponential mixtures \cite{anas1998urban}, where population density is written as previously in Eq.~\ref{eq:kernelmixture}. Parameters for this model are the number of kernels $n^{(E)}$, the rank-size hierarchy $\alpha^{(E)}$, the size of the largest kernel $r_0^{E}$. Contrary to previous models in which total population had an influence by controlling the speed of growth, density can here be rescaled arbitrarily (morphological indicators used are not changed through rescaling, see below), and we set the maximal density for one kernel to one.

% null model: white noise? (for robustness of indicators)

\subsection*{Measures of urban form}

% already computed for dynamical calib for world urban areas -> use same measures. todo future work: benchmark measures world areas: landscape eco, etc.

Quantitative measures of urban form are multiple and depend on the scale considered \cite{zhang2005metrics}. \cite{raimbault2019generating} for example introduces measures for buildings at the district scale. The field of Landscape Ecology has its own metrics similar to urban form measures \cite{10.1371/journal.pone.0225734}. For the scale we consider and considering population distribution only, metrics of urban form have been proposed for example to quantify sprawl \cite{tsai2005quantifying}. These can be related to fractal approaches to urban form \cite{chen2011derivation}. The effective dimension when applied to real cities is reasonably low \cite{schwarz2010urban}, and a few complementary indicators can be used. We thus follow \cite{10.1371/journal.pone.0203516} and consider urban form measures which are: (i) Moran index to capture spatial autocorrelation and the existence of centers; (ii) average distance between individuals which captures a level of aggregation; (iii) distribution entropy (aspatial) to capture the uniformity of the distribution; and (iv) rank-size slope which captures the hierarchy of population distribution.

\section*{Results}

\subsection*{Implementation}

The models are implemented in \texttt{scala} and integrated into the \texttt{spatialdata} library for spatial sensitivity analysis \cite{raimbault2020scala}. The library is bundled as an OpenMOLE plugin for the numerical experiments. OpenMOLE is an open source software for model exploration and validation \cite{reuillon2013openmole} combining model embedding with state-of-the-art exploration methods (including for example sensitivity analysis, design of experiments, calibration with genetic algorithms) and a transparent distribution of computations on high performance computing environments. In our case, we use its workflow system and an integrated algorithm to determine the feasible space of models.

\subsection*{Empirical data}

%%%%%%%%%%%%%%
\begin{figure}[!h]
	\includegraphics[width=\linewidth]{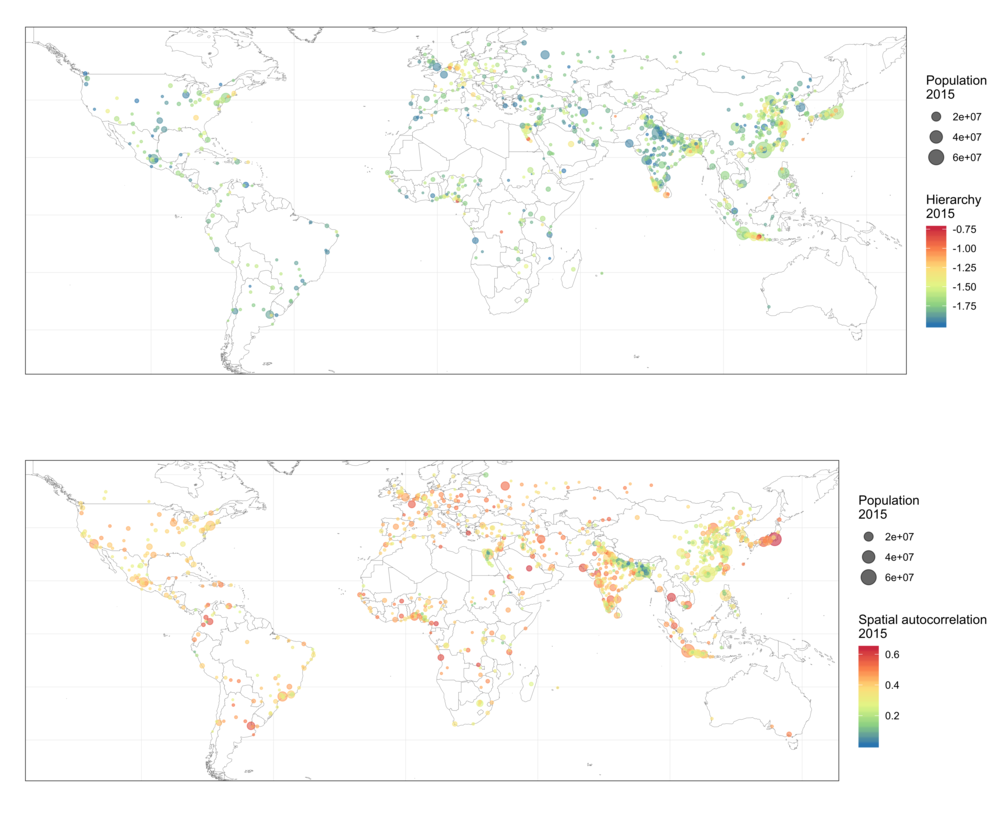}
	\caption{{\bf Maps of urban morphology indicators for worldwide urban areas.} (Top) Rank-size hierarchy; (Bottom) Moran spatial autocorrelation index. Area of circles gives population.\label{fig:fig1}}
\end{figure}
%%%%%%%%%%%%%%

%%%%%%%%%%%%%%
\begin{figure}[!h]
	\includegraphics[width=\linewidth]{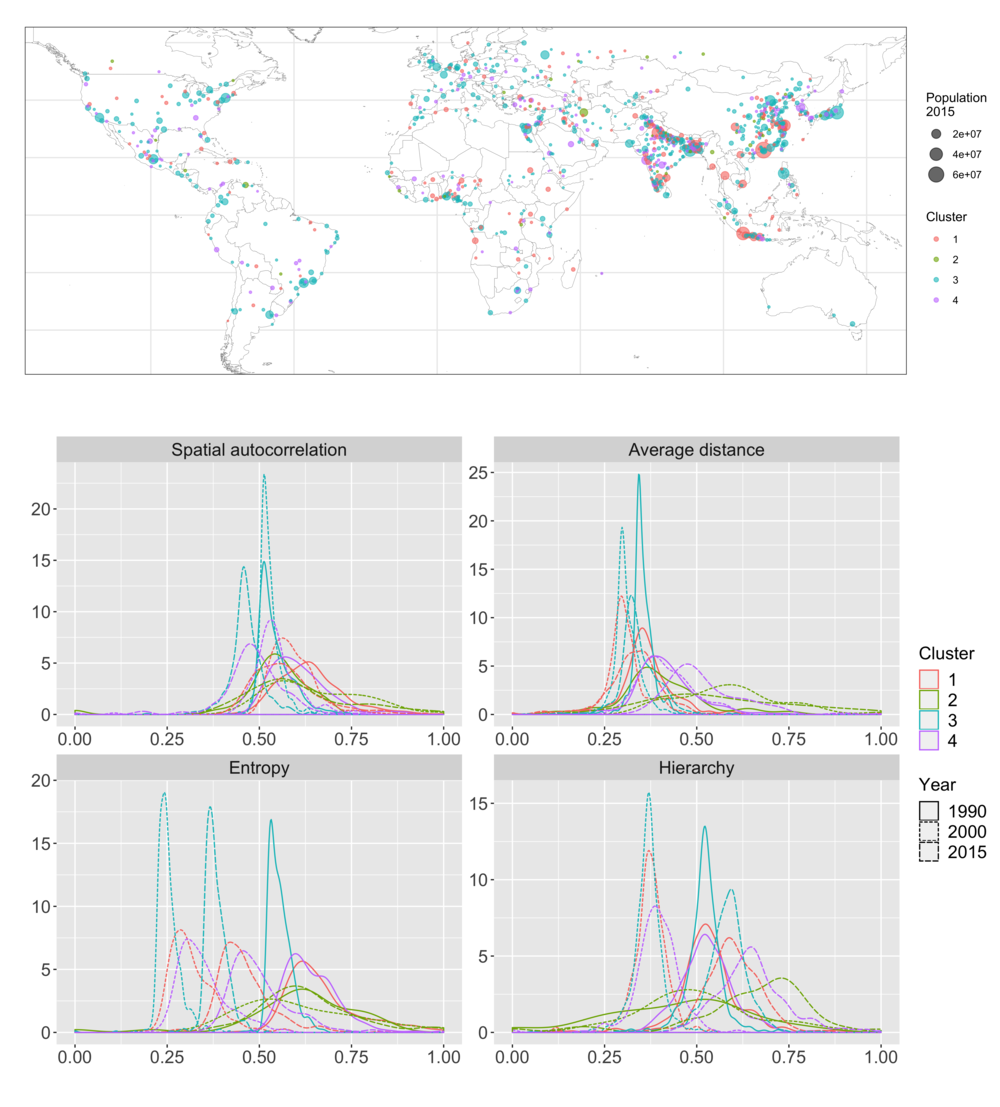}
	\caption{{\bf Urban form typology. Using an unsupervised k-means clustering algorithm, we find clusters among urban areas.} (Top) Map of cluster belonging; (Bottom) Statistical distribution of indicator values, within each cluster and in time for all dates in the GHSL dataset.\label{fig:fig2}}
\end{figure}
%%%%%%%%%%%%%%

We compare model simulation results to urban form measures computed on worldwide urban areas. We use therefore the Global Human Settlement Layer database (GHSL), which provides an exhaustive worldwide population raster with a 1km resolution \cite{melchiorri2018unveiling}. \cite{Raimbault_2020} has shown the relevance of using this database for worldwide simulation models. It is available at four dates from 1975 onwards, but as not all models are dynamical we use the most recent population configuration only (2015). The database provides a layer of urban areas, within which we extract for the 1000 largest areas in terms of population a covering window ($\pm 25\%$ of extent on each side) from the population raster (note that some windows may be overlapping, as in the case of Hong-Kong which is separated from the main cluster of the Pearl River Delta mega-city region). Morphological indicators are computed on these extracted areas.

We show in Fig.~\ref{fig:fig1} maps of indicators. More particularly, we map the rank-size hierarchy and Moran spatial autocorrelation which have meaningful geographical variations. Rank-size hierarchy will tell if the metropolitan area is strongly dominated by one center or if it more balanced. We retrieve the fact that in Europe, Paris and London are known for such a strong monocentricity, compared to cities in Germany for example. Similarly, mega-city regions in East Asia (Pearl River delta, Yangtze River delta, Beijing-Tianjin) are more polycentric and thus balanced than Wuhan or Seoul. Regarding spatial autocorrelation, we also observe a strong variation in East Asia (Tokyo compared to Chinese megacities for example), and within India (agglomerations in the Gange plain making a cluster of non-correlated, thus highly sprawled areas).

The areas can be clustered following a non-supervised approach. We proceed to a k-means clustering on normalized indicator values, and find $k=4$ clusters as meaningful regarding the derivative of within-cluster variance. A map of cluster belonging and cluster profiles are shown in Fig.~\ref{fig:fig2}. We retrieve the variation within East Asia (Tokyo, Seoul and Shanghai being each in a different cluster) but less in Europe. The main cluster (light blue) corresponds to strongly monocentric urban areas. We see in density distributions of indicator values that clusters have clearly different profiles, which correspond to different typologies of urban morphology that the models try to approximate. For example, cluster 3 (light blue) and 1 (red) have the same level of hierarchy, but the latter has a much higher autocorrelation and entropy, and corresponds thus to more polycentric configurations.

%\subsection*{Behavior of models}
% also saltelli?

%%%%%%%%%%%%%%
\begin{figure}[!h]
	\includegraphics[width=\linewidth]{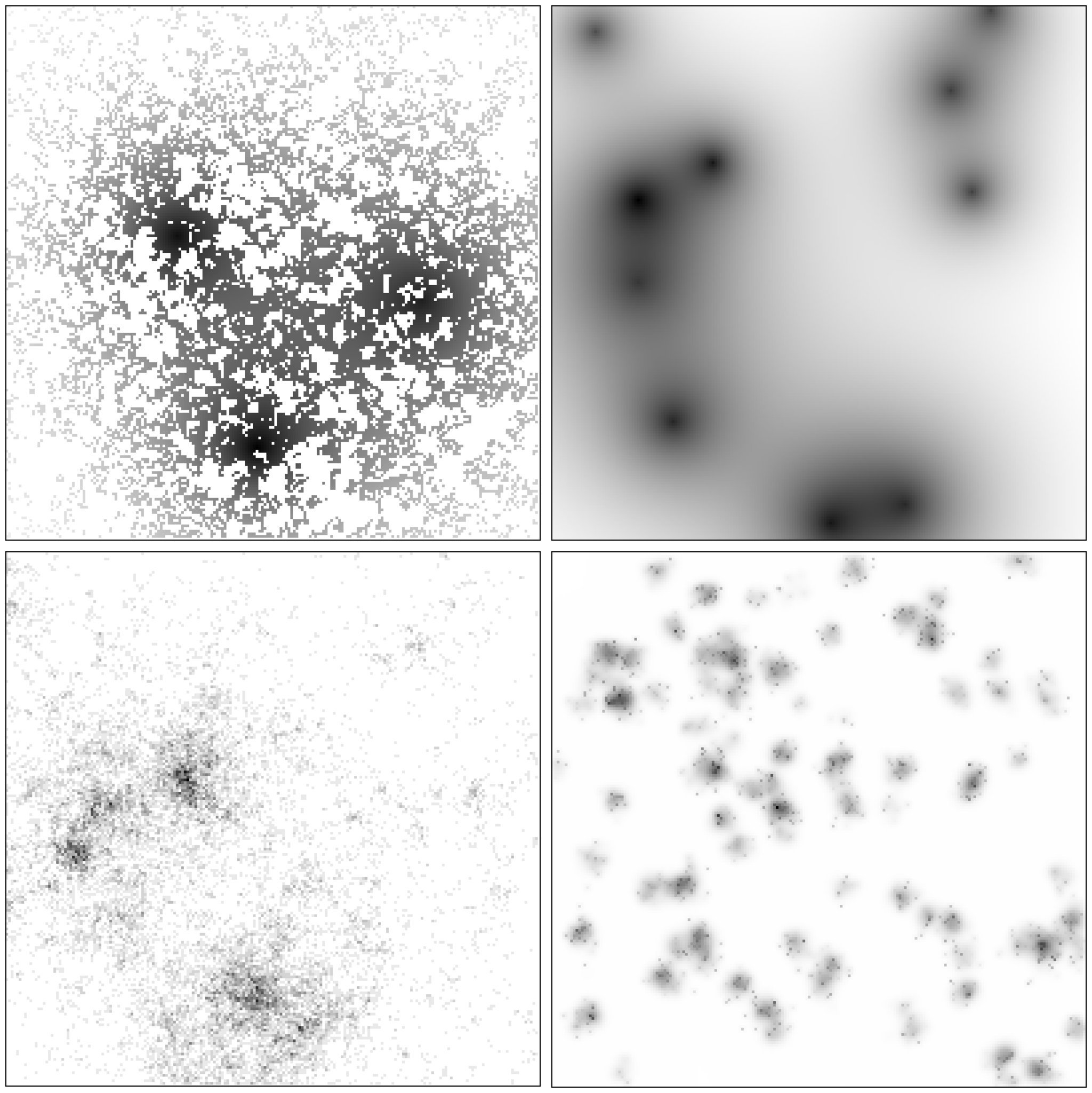}
	\caption{{\bf Examples of generated urban forms for a square world of width $W = 50$.} From Left to Right and Top to Bottom: (i) Correlated percolation model for $r_0^{(C)} = 50$, $\alpha^{(C)} = 0.4$, $n^{(C)} = 3$; (ii) Exponential mixture model for $n^{(E)} = 10$, $r_0^{(E)} = 40$, $\alpha^{(E)} = 1$; (iii) Gravity model for $g^{(G)} = 0.3$, $\gamma_D^{(G)} = 2.5$, $\gamma_P^{(G)} = 0.5$, $P_0^{(G)} = 3$, $P_M^{(G)} = 20000$; (iv) Reaction-diffusion model for $\alpha^{(R)} = 0.8$, $\beta^{(R)} = 0.2$, $d^{(R)} = 1$, $N^{(R)} = 100$, $P_M^{(R)} = 5000$.\label{fig:fig3}}
\end{figure}
%%%%%%%%%%%%%%

\subsection*{Generated urban forms}

% examples for each model

We show in Fig.~\ref{fig:fig3} examples of generated urban forms for each model included in this study. Visually, these urban form look rather different. To what extent they are statistically distant for the morphology indicators can only be determined by systematic experiments. The gravity and percolation results look similar, although at a slightly different scale. This particular configuration of the reaction-diffusion model corresponds more to a rural or peri-urban configuration, while the exponential mixture is fuzzy and would resemble a blurred polycentric urban configuration.

%%%%%%%%%%%%%%
\begin{figure}[!h]
	\includegraphics[width=\linewidth]{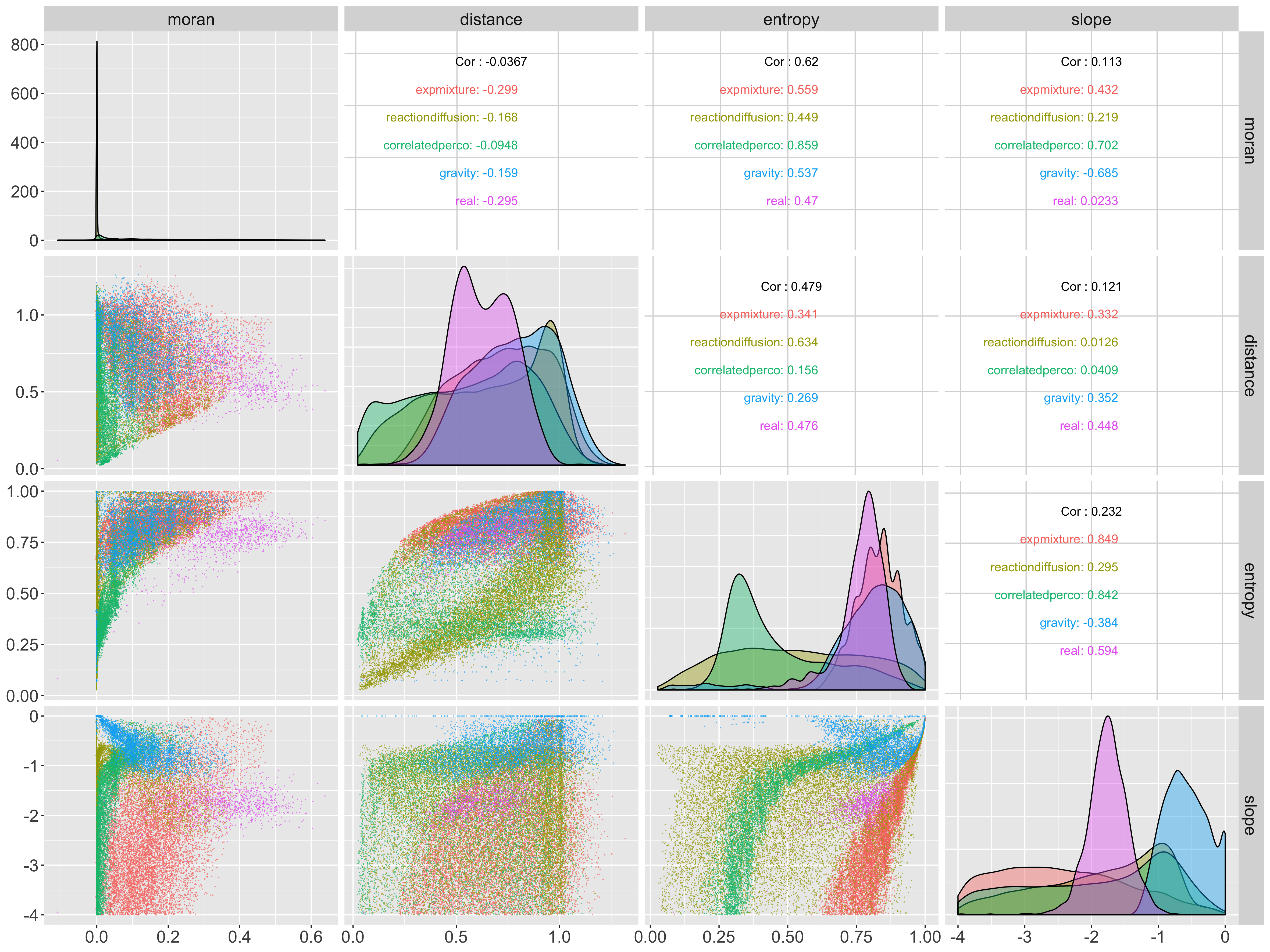}
	\caption{{\bf Feasible morphological space for each simulation model.} The lower part of the plot matrix gives a scatterplot between each indicator. We also plot real urban areas in purple. The diagonal gives statistical distribution, while the correlation matrix is given in the upper part, conditionally to each model.\label{fig:fig4}}
\end{figure}
%%%%%%%%%%%%%%

\subsection*{Feasible morphological spaces}

We now turn to the main experiment of this paper: using a diversity search algorithm to determine the full feasible morphological space for each model. Therefore, we use the Pattern Space Exploration (PSE) algorithm \cite{cherel2015beyond} embedded in OpenMOLE. Diversity search was introduced in the field of artificial life as genetic algorithms with the aim to maximize diversity of the population \cite{lehman2008exploiting}. In the case of the PSE algorithm, a novelty criteria leads the search towards new regions of the indicator space, and results are stored in a hitmap. When running long enough, convergence in terms of number of solutions found is generally reached, and one can consider the final population as the feasible space of the model. We run here the algorithm up to 100,000 generations for each model, with grid of step 0.05 in the indicator space. Convergence was reached separately for each model.

We show in Fig~\ref{fig:fig4} a scatterplot of the final population in the morphological space. First of all, we observe that correlation are very different for each model (for example opposite value for gravity and reaction-diffusion between entropy and slope), confirming that the way urban form is produced is very different although the final form may be the same. Then, we observe that the point clouds slightly intersect but also have their own proper space that which no other model can reach. For example, in the entropy-slope space, the reaction-diffusion model is very flexible and covers a large part of space, while gravity is restricted to a small region and correlated percolation and exponential mixture exhibit much less variability with a strong correlation between the two dimensions. The models cover a similar region for the Moran-entropy space, but all fail to capture an important part of real points (in purple), corresponding to configurations with a high Moran index but a relatively low entropy. Models systematically associate a high moran with a high entropy. In other dimensions, real points are reasonably covered. Thus, we can conclude that (i) the different models are complementary in terms of urban forms produced and (ii) that a part of real configurations are approached by some of the models, but other can not be reproduced.

%%%%%%%%%%%%%%
\begin{figure}[!h]
	\begin{center}
	\includegraphics[width=0.6\linewidth]{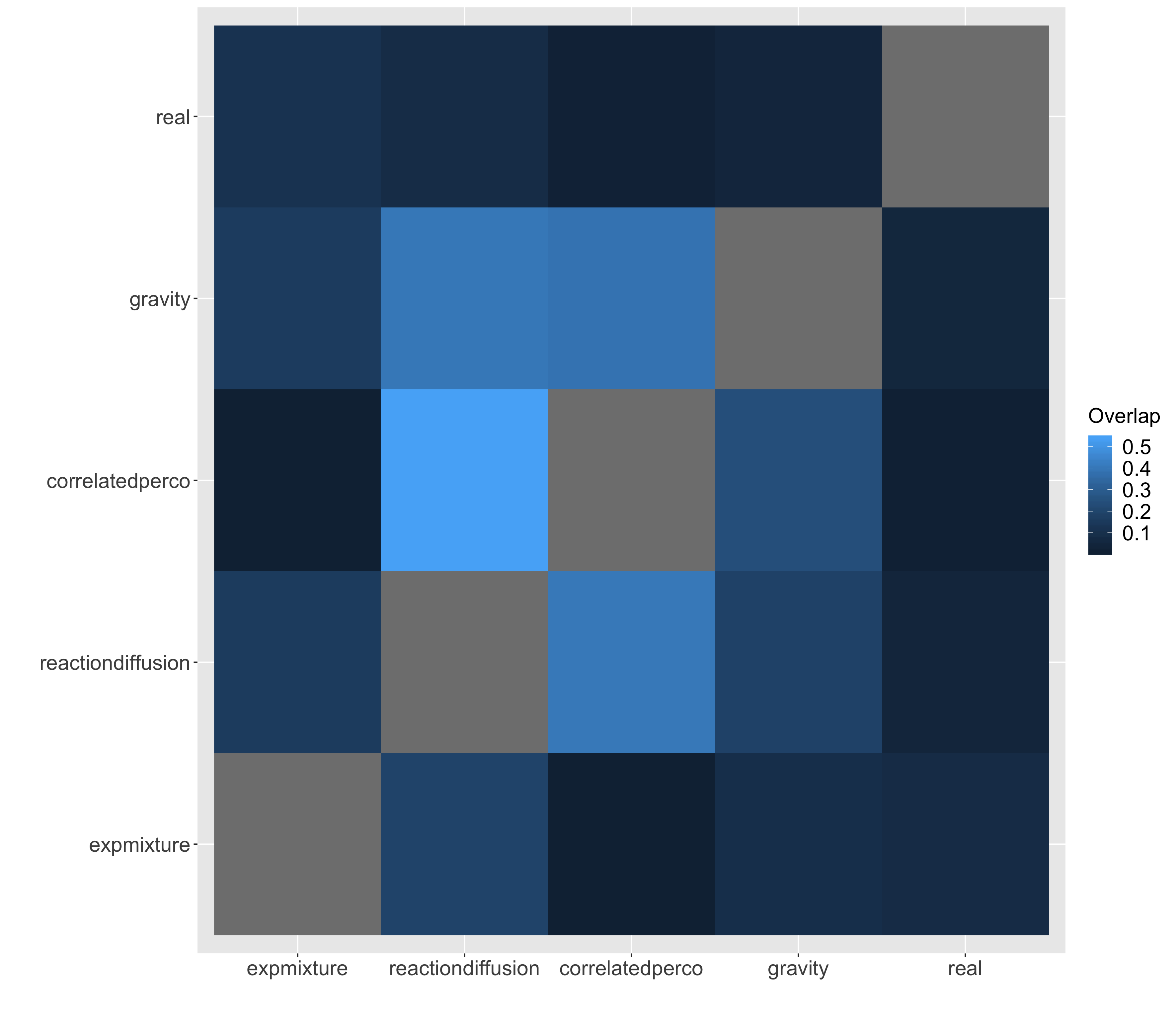}
	\end{center}
	\caption{{\bf Overlap between point cloud hypervolumes.}\label{fig:fig5}}
\end{figure}
%%%%%%%%%%%%%%

To better quantify how each model are similar and how they approach real configurations, we compute the hypervolume (in the four dimensional indicator space) corresponding to each point cloud and the intersections between these hypervolumes. We use the \texttt{hypervolume} \texttt{R} package \cite{hypervolume} with a gaussian kernel density estimation with adaptable bandwidth. Then, we compute for each couple of model the ratio between the intersection of hypervolumes and the second model. The non-symmetric relative overlap matrix is plotted in Fig.~\ref{fig:fig5} (diagonal was removed for a better readability). We find that the closer model are the correlated percolation and reaction diffusion, with close to 50\% overlap. Then the gravity point cloud is mostly contained by reaction-diffusion and correlated percolation, but not the other way around. Thus models produce similar points in the parameter space, but most of their output cloud is original. Finally, we find regarding the proximity to real configurations that the best model is the exponential mixture. At the price of producing blurry urban forms, the flexibility therein is the best to approach the best existing configurations. Then comes the reaction-diffusion model, and the least flexible to approach real values is the correlated percolation model.

\section*{Discussion}

% results, implications

We have shown that different simple urban morphogenesis models are strongly complementary in the morphological space, confirming that a complementarity of processes also leads to a complementarity in patterns produced. This result rejoins the results obtained by \cite{raimbault2018multi} in the case of transportation network generative models at a similar mesoscopic scale, the results of \cite{raimbault2019generating} for generating building configurations, and the results of \cite{raimbault2018systematic} in the case of dynamical models for systems of cities. Both found a complementarity of approach including different types of processes. We argue that this is further evidence of the multi-dimensionality of urban systems and for a necessity of a plurality of urban models to capture both diverse processes but also outcomes.

% future dev

% dynamical calib

This work is a first step towards a systematic benchmark of simple urban morphogenesis models. Future work should include the search for an explanation of the unreachable real urban configuration, and possibly alternative models approaching these. Other urban form indicators should also be tested. Finally, dynamical calibration of models remains an open question, investigated in the case of the reaction-diffusion model by \cite{raimbault:halshs-02406539}. An issue is that some models such as the correlated percolation model are not dynamical and should thus be adapted to be calibrated between successive points in time.

\section*{Conclusion}

We have implemented and systematically compared four very different simple models for urban morphogenesis, including reaction-diffusion, correlated percolation, exponential mixture and a gravitational aggregation model. We applied a diversity search algorithm to obtain the feasible morphological space. The results confirm a complementarity between the models and the relevance of a plurality in urban modeling approaches.

%\section*{Supporting information}

%\section*{Acknowledgments}

\nolinenumbers

% Either type in your references using
% \begin{thebibliography}{}
% Text
% \end{thebibliography}
%
% or
%
% Compile your BiBTeX database using our plos2015.bst
% style file and paste the contents of your .bbl file
% here. See http://journals.plos.org/plosone/s/latex for 


\begin{thebibliography}{10}

\bibitem{batty2018inventing}
Batty M.
\newblock Inventing future cities.
\newblock MIT press; 2018.

\bibitem{williams2000achieving}
Williams K, Burton E, Jenks M.
\newblock Achieving sustainable urban form: an introduction.
\newblock Achieving sustainable urban form. 2000;2000:1--5.

\bibitem{le2012urban}
Le~N{\'e}chet F.
\newblock Urban spatial structure, daily mobility and energy consumption: a
  study of 34 european cities.
\newblock Cybergeo: European Journal of Geography. 2012;.

\bibitem{viguie2012trade}
Vigui{\'e} V, Hallegatte S.
\newblock Trade-offs and synergies in urban climate policies.
\newblock Nature Climate Change. 2012;2(5):334--337.

\bibitem{wegener2004land}
Wegener M, F{\"u}rst F.
\newblock Land-use transport interaction: state of the art.
\newblock Available at SSRN 1434678. 2004;.

\bibitem{milton2019accelerating}
Milton R, Roumpani F.
\newblock Accelerating Urban Modelling Algorithms with Artificial Intelligence.
\newblock In: Proceedings of the 5th International Conference on Geographical
  Information Systems Theory, Applications and Management. vol.~1. INSTICC;
  2019. p. 105--116.

\bibitem{batty1997cellular}
Batty M.
\newblock Cellular automata and urban form: a primer.
\newblock Journal of the American Planning Association. 1997;63(2):266--274.

\bibitem{makse1998modeling}
Makse HA, Andrade JS, Batty M, Havlin S, Stanley HE, et~al.
\newblock Modeling urban growth patterns with correlated percolation.
\newblock Physical Review E. 1998;58(6):7054.

\bibitem{batty2007cities}
Batty M.
\newblock Cities and complexity: understanding cities with cellular automata,
  agent-based models, and fractals.
\newblock The MIT press; 2007.

\bibitem{murcio2015urban}
Murcio R, Morphet R, Gershenson C, Batty M.
\newblock Urban transfer entropy across scales.
\newblock PLoS One. 2015;10(7):e0133780.

\bibitem{batty1989urban}
Batty M, Longley P, Fotheringham S.
\newblock Urban growth and form: scaling, fractal geometry, and
  diffusion-limited aggregation.
\newblock Environment and planning A. 1989;21(11):1447--1472.

\bibitem{murcio2009colored}
Murcio R, Rodr{\'\i}guez-Romo S.
\newblock Colored diffusion-limited aggregation for urban migration.
\newblock Physica A: Statistical Mechanics and its Applications.
  2009;388(13):2689--2698.

\bibitem{murcio2013second}
Murcio R, Sosa-Herrera A, Rodriguez-Romo S.
\newblock Second-order metropolitan urban phase transitions.
\newblock Chaos, Solitons \& Fractals. 2013;48:22--31.

\bibitem{10.1371/journal.pone.0203516}
Raimbault J.
\newblock Calibration of a density-based model of urban morphogenesis.
\newblock PLOS ONE. 2018;13(9):1--18.
\newblock doi:{10.1371/journal.pone.0203516}.

\bibitem{li2019singularity}
Li Y, Rybski D, Kropp JP.
\newblock Singularity cities.
\newblock Environment and Planning B: Urban Analytics and City Science. 2019;
  p. 2399808319843534.

\bibitem{pumain2020conclusion}
Pumain D, Raimbault J.
\newblock Conclusion: Perspectives on urban theories.
\newblock In: Theories and Models of Urbanization. Springer; 2020. p. 303--330.

\bibitem{fotheringham1989spatial}
Fotheringham AS, O'Kelly ME.
\newblock Spatial interaction models: formulations and applications. vol.~1.
\newblock Kluwer Academic Publishers Dordrecht; 1989.

\bibitem{rybski2013distance}
Rybski D, Ros AGC, Kropp JP.
\newblock Distance-weighted city growth.
\newblock Physical Review E. 2013;87(4):042114.

\bibitem{turing1990chemical}
Turing AM.
\newblock The chemical basis of morphogenesis.
\newblock Bulletin of mathematical biology. 1990;52(1-2):153--197.

\bibitem{raimbault2018co}
Raimbault J.
\newblock Co-evolution and morphogenetic systems.
\newblock arXiv preprint arXiv:180311457. 2018;.

\bibitem{bonin2014modelisation}
Bonin O, Hubert JP.
\newblock Mod{\'e}lisation morphog{\'e}n{\'e}tique de moyen terme des villes:
  une sch{\'e}matisation du mod{\`e}le th{\'e}orique de Ritchot et Desmarais
  dans le cadre du mod{\`e}le standard de l'{\'e}conomie urbaine.
\newblock Revue dEconomie Regionale Urbaine. 2014;(3):471--497.

\bibitem{parish2001procedural}
Parish YI, M{\"u}ller P.
\newblock Procedural modeling of cities.
\newblock In: Proceedings of the 28th annual conference on Computer graphics
  and interactive techniques; 2001. p. 301--308.

\bibitem{makse1996method}
Makse HA, Havlin S, Schwartz M, Stanley HE.
\newblock Method for generating long-range correlations for large systems.
\newblock Physical Review E. 1996;53(5):5445.

\bibitem{anas1998urban}
Anas A, Arnott R, Small KA.
\newblock Urban spatial structure.
\newblock Journal of economic literature. 1998;36(3):1426--1464.

\bibitem{zhang2005metrics}
Zhang M, Kukadia N.
\newblock Metrics of urban form and the modifiable areal unit problem.
\newblock Transportation Research Record. 2005;1902(1):71--79.

\bibitem{raimbault2019generating}
Raimbault J, Perret J.
\newblock Generating urban morphologies at large scales.
\newblock In: Artificial Life Conference Proceedings. MIT Press; 2019. p.
  179--186.

\bibitem{10.1371/journal.pone.0225734}
Bosch M.
\newblock PyLandStats: An open-source Pythonic library to compute landscape
  metrics.
\newblock PLOS ONE. 2019;14(12):1--19.
\newblock doi:{10.1371/journal.pone.0225734}.

\bibitem{tsai2005quantifying}
Tsai YH.
\newblock Quantifying urban form: compactness versus' sprawl'.
\newblock Urban studies. 2005;42(1):141--161.

\bibitem{chen2011derivation}
Chen Y.
\newblock Derivation of the functional relations between fractal dimension of
  and shape indices of urban form.
\newblock Computers, Environment and Urban Systems. 2011;35(6):442--451.

\bibitem{schwarz2010urban}
Schwarz N.
\newblock Urban form revisited—Selecting indicators for characterising
  European cities.
\newblock Landscape and urban planning. 2010;96(1):29--47.

\bibitem{raimbault2020scala}
Raimbault J, Perret J, Reuillon R.
\newblock A scala library for spatial sensitivity analysis.
\newblock arXiv preprint arXiv:200710667. 2020;.

\bibitem{reuillon2013openmole}
Reuillon R, Leclaire M, Rey-Coyrehourcq S.
\newblock OpenMOLE, a workflow engine specifically tailored for the distributed
  exploration of simulation models.
\newblock Future Generation Computer Systems. 2013;29(8):1981--1990.

\bibitem{melchiorri2018unveiling}
Melchiorri M, Florczyk AJ, Freire S, Schiavina M, Pesaresi M, Kemper T.
\newblock Unveiling 25 years of planetary urbanization with remote sensing:
  Perspectives from the global human settlement layer.
\newblock Remote Sensing. 2018;10(5):768.

\bibitem{Raimbault_2020}
Raimbault J, Denis E, Pumain D.
\newblock Empowering Urban Governance through Urban Science: Multi-Scale
  Dynamics of Urban Systems Worldwide.
\newblock Sustainability. 2020;12(15):5954.
\newblock doi:{10.3390/su12155954}.

\bibitem{cherel2015beyond}
Ch{\'e}rel G, Cottineau C, Reuillon R.
\newblock Beyond corroboration: Strengthening model validation by looking for
  unexpected patterns.
\newblock PloS one. 2015;10(9):e0138212.

\bibitem{lehman2008exploiting}
Lehman J, Stanley KO.
\newblock Exploiting open-endedness to solve problems through the search for
  novelty.
\newblock In: ALIFE; 2008. p. 329--336.

\bibitem{hypervolume}
Blonder B, with contributions~from David J~Harris. hypervolume: High
  Dimensional Geometry and Set Operations Using Kernel Density Estimation,
  Support Vector Machines, and Convex Hulls; 2019.
\newblock Available from: \url{https://CRAN.R-project.org/package=hypervolume}.

\bibitem{raimbault2018multi}
Raimbault J.
\newblock Multi-modeling the morphogenesis of transportation networks.
\newblock In: Artificial Life Conference Proceedings. MIT Press; 2018. p.
  382--383.

\bibitem{raimbault2018systematic}
Raimbault J.
\newblock A systematic comparison of interaction models for systems of cities.
\newblock In: Conference on Complex Systems 2018; 2018.

\bibitem{raimbault:halshs-02406539}
Raimbault J.
\newblock {Worldwide estimation of parameters for a simple reaction-diffusion
  model of urban growth}.
\newblock In: {International Land-use Symposium 2019}. Paris, France;
  2019.Available from:
  \url{https://halshs.archives-ouvertes.fr/halshs-02406539}.

\end{thebibliography}
\end{document}